\documentclass[]{aa}

\usepackage{graphicx}
\usepackage[varg]{txfonts}
\usepackage{natbib}

\begin{document}

\title {Case study on the identification and classification of small-scale flow patterns in flaring active region}

\author{E. Philishvili \inst{1,2}, B.M. Shergelashvili \inst{3,2,4}, S. Buitendag \inst{5}, J. Raes \inst{1}, S. Poedts \inst{1,6}, M.L. Khodachenko \inst{3,7,8}}

\institute{ Centre for Mathematical Plasma Astrophysics/Department of Mathematics, Celestijnenlaan 200B, 3001, Leuven, Belgium\\
 \and
Centre for Computational Helio Studies, Ilia State University,
Cholokashvili Ave. 3/5, 0162 Tbilisi, Georgia\\
  \and
Space Research Institute, Austrian Academy of Sciences, Schmiedlstrasse 6, 8042 Graz, Austria\\
 \and
Evgeni Kharadze Georgian National Astrophysical Observatory, Abastumani, 0301 Adigeni Municipality, Georgia\\
 \and
Department of Statistics and Actuarial Science, Stellenbosch University, Victoria Street, 7600, Stellenbosch, South Africa \\
 \and
Institute of Physics, University of Maria Curie-Sk{\l}odowska, PL-20-031 Lublin, Poland\\
  \and
Lomonosov Moscow State University, Skobeltsyn Institute of Nuclear Physics (MSU SINP)  Leninskie Gory, 119992, Moscow, Russia\\
  \and
Institute of Astronomy, Russian Academy of Sciences, Moscow 119017, Russia}

\abstract
{We propose a novel methodology to identity flows in the solar atmosphere and classify their velocities as either supersonic, subsonic, or sonic.  }
{The proposed methodology consists of three parts. First, an algorithm is applied to the Solar Dynamics Observatory (SDO) image data to locate and track flows, resulting in the trajectory of each flow over time. Thereafter, the differential emission measure inversion method is applied to six Atmospheric Imaging Assembly (AIA)  channels along the trajectory of each flow in order to estimate its background temperature and sound speed. Finally, we classify each flow as supersonic, subsonic, or sonic by performing simultaneous hypothesis tests on whether the velocity bounds of the flow are larger, smaller, or equal to the background sound speed.}
{The proposed methodology was applied to the SDO image data from the 171$\AA$ spectral line for the date  6 March 2012 from 12:22:00 to 12:35:00 and again for the date 9 March 2012 from 03:00:00 to 03:24:00. Eighteen plasma flows were detected, 11 of which were classified as supersonic, 3 as subsonic, and 3 as sonic at a $70\%$ level of significance. Out of all these cases, 2 flows cannot be strictly ascribed to one of the respective categories as they change from the subsonic state to supersonic and vice versa. We labeled them as a subclass of transonic flows.}
{The proposed methodology provides an automatic and scalable solution to identify small-scale flows and to classify their velocities as either supersonic, subsonic, or sonic. It can be used to characterize the physical properties of the solar atmosphere. }
{We identified and classified small-scale flow patterns in flaring loops. The results show that the flows can be classified into four classes: sub-, super-, trans-sonic, and sonic. The flows occur in the complex structure of the active region magnetic loops. The detected flows from AIA images can be analyzed in combination with the other high-resolution observational data, such as Hi-C 2.1 data, and be used for the development of theories describing the physical conditions responsible for the formation of flow patterns.}


\titlerunning{Flow patterns}

\authorrunning{Philishvili et al.}

\maketitle

\section{Introduction}

The solar corona hosts many different eruptive and explosive events, which are manifest in the various flow patterns found in these complex magnetic loop systems. In the past few decades, high temporal and spatial resolution imaging telescopes (Solar Dynamics Observatory (SDO), Solar Terrestrial Relations Observatory (STEREO), Solar and Heliospheric Observatory (SOHO), etc.)  have made it possible to observe the intensity changes along the magnetic loops with characteristic spacial scales of at least tens of resolution pixels (for instance, SDO has a resolution of $1''.2$ and SOHO of $5''$). This has led to significant efforts to study the physical properties of the intermediate- and large-scale plasma flows as compared to the visible size of the magnetic structure of interest (e.g., active regions, filaments, coronal bright points)  induced during solar flares, including micro- or nano-flares. The main purpose of these efforts is to understand both the direct and indirect effects of energy releases into the ambient solar atmosphere \citep[for an example of such an investigation and a comprehensive outline of related literature, see ][]{Kamio2011}. These flow phenomena help to establish the cause-effect relationship between properties of the magnetic structures and the plasma flows, which in turn determine the related physical conditions of the solar atmosphere (see \citet{Srivastava2019} and \citet{Reale2014}).

However, the observations of smaller scale flows are rather puzzling as most of these flows are of the order of a few times the spatial resolution limit, provided by the instruments on the abovementioned space missions. Recently, \citet{Williams2020} reported the discovery of a large population of so-called coronal strands in active regions made with the High-Resolution Coronal Imager (Hi-C 2.1, with $0''.129\;$pixel$^{-1}$ resolution, see \citet{Rachmeler2019}), which is the advanced version of the original Hi-C NASA sounding rocket experiment performed in 2012 \citep{Kobayashi2014}.  The existence of these strands reveals that large-scale magnetic objects like active regions have fine small-scale structures. In order to understand the energetics of the flows in the context of the overall energy budget of an active region, it is important to study the nature of the motion of the intensity blobs and to classify them in terms of their characteristic dynamic parameters. For instance, it is important to know which of them are engaged in oscillatory or aperiodic flow-type motions. An example case study of such oscillations is discussed in \citet{Philishvili2017}, where oscillatory flows in the active region AR11429 are interpreted as standing longitudinal (acoustic) modes with periods within the range of $116-147\;$sec. The nature of these oscillations drastically differs from those with long periods ($>1\;$hour) reported by \citet{Dumbadze2017} and references therein. As to the aperiodic motions of the observed intensity blobs in the same active region (AR11429), these are the main objective of the present study. In particular, we aim to formulate a mathematically rigorous identification and statistical classification method using SDO/AIA 171$\AA$ for flows in the small-scale strands of similar spatial scales as those observed with the higher resolution instruments \citep{Williams2020}.
Our proposed method provides a set of identified aperiodic flow trajectories together with their estimated velocity bounds (taking into account corrections in pixel distances due to the projection deformation of the solar spherical surface in flat CCD images) and classification as either subsonic, sonic, or supersonic. We will do this by estimating the local background sound speeds by means of the differential emission measure (DEM) inversion algorithm of \citet{Plowman2013}.

\begin{figure*}[h!]
  \includegraphics[width=60mm, trim={5mm 0 5mm 0}, clip]{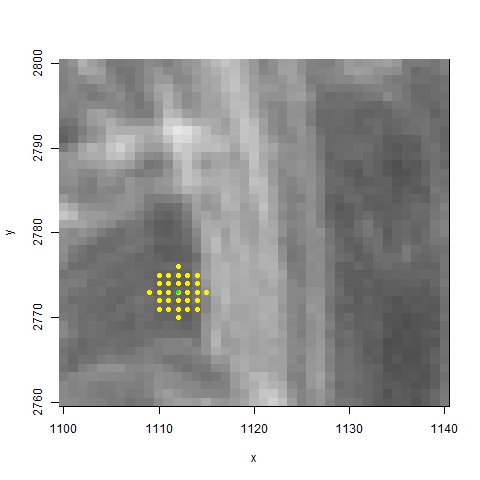}
    \includegraphics[width=60mm, trim={5mm 0 5mm 0}, clip]{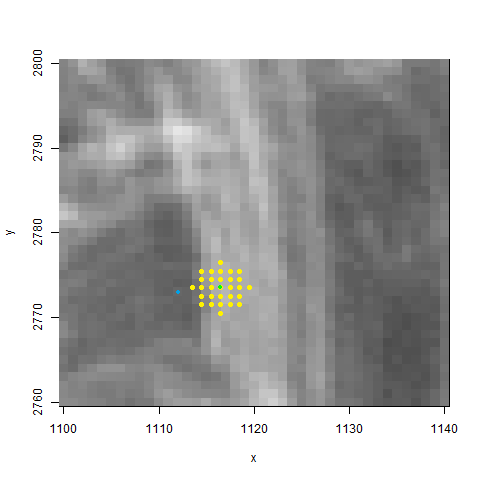}
      \includegraphics[width=60mm, trim={5mm 0 5mm 0}, clip]{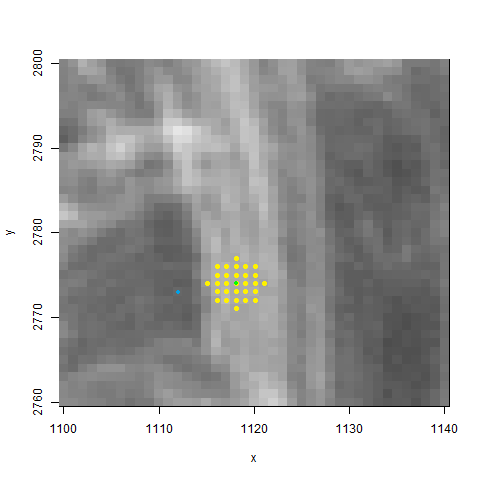}
  \caption{Illustration of a centroid with a radius of three pixels converging over iterations 1 (left), 6 (center), and 12 (right). The blue point indicates the start point of the centroid, and the green point indicates the current centroid estimate.}
  \label{converge}
\end{figure*}

The investigation of flows in the solar corona with a variety of spatiotemporal scales is an important task as they carry information about their acceleration sources, which are related to the thermal and mechanical energy release during flows and sporadic energy release events as mentioned above. Dynamic mass flows in coronal magnetic loops are a typical attribute of a rapid energy deposition in the atmosphere. These flows affect the coronal heating rate by evaporating \citep{Acton1982, Hirayama1974, Neupert1968} and condensing chromospheric plasma \citep{Canfield1990, Fisher1987} into the corona. \citet{Y.Li2015} observed the chromospheric evaporation process in flaring loops and \citet{Nitta2012} found similar upward flow motions with $100-500\;$km/s velocities associated with small transient brightening. These events can be separated into two different types \citep{Y.Li2015}: subsonic flows (gentle evaporation), which are characterized by tens of kilometers per second velocities, and supersonic flows (explosive evaporation) which are characterized by velocities of hundreds of kilometers per second. The rapid acceleration of the plasma can be energetically supported by thermal \citep{Shergelashvili2007} and magnetohydrodynamic \citep{Shergelashvili2006} instability processes. In coronal hole bright points \citep{Bagashvili2018} (representing small-scale, active-region-like structures) and in active regions \citep{Joshi2020},  quasi-oscillatory variability has been discovered to precede jet outflows, which can be interpreted as indications of the presence of instability processes. Furthermore, upflows are observed at the edges of active regions \citep{Boutry2012}. All these types of 'intermediate scale' flow patterns are proposed to contribute to the acceleration of the large-scale solar wind in the inner heliosphere (see \citet{Brooks2010}). 
It has been shown that the existence of magnetic reconnection sources in the corona can lead to extremely rapid acceleration, even forming discontinuous patterns of the solar wind \citep{Shergelashvili2020}. 
\citet{Shergelashvili2012} and \citet{Vainio2003} discuss a causal chain in which the energy of intermediate-scale flows can be dissipated into heat and wave turbulence energy  at the source region where the fast and slow solar wind patterns are formed. While the latter determine the space weather conditions in the interplanetary space and shape the entire heliosphere, it is plausible that the small-scale flows are also members of this chain. 
Small-scale flows are abundant in active regions and other magnetic structures, and represent carriers of energy for the formation of the multi-scale structures of magnetic objects, their distortions, and instabilities in the solar corona. Therefore, a thorough understanding of the dynamical properties of those small-scale flows, with the future perspective to upgrade towards full-scale statistical, machine, or deep learning models, is a crucially important problem. 

\begin{figure*}[h!]
  \includegraphics[width=60mm, trim={5mm 0 5mm 0}, clip]{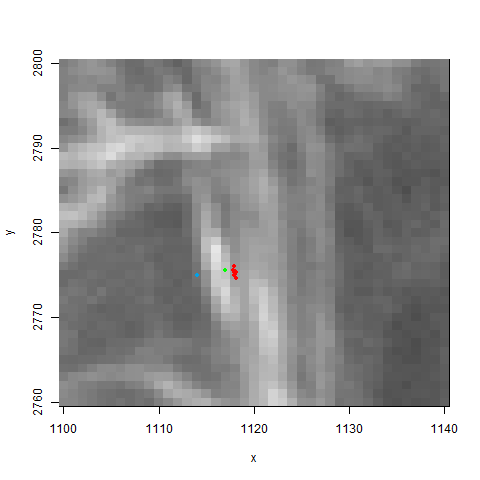}
    \includegraphics[width=60mm, trim={5mm 0 5mm 0}, clip]{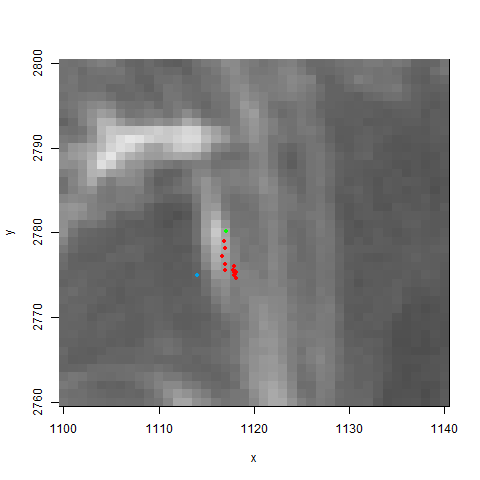}
      \includegraphics[width=60mm, trim={5mm 0 5mm 0}, clip]{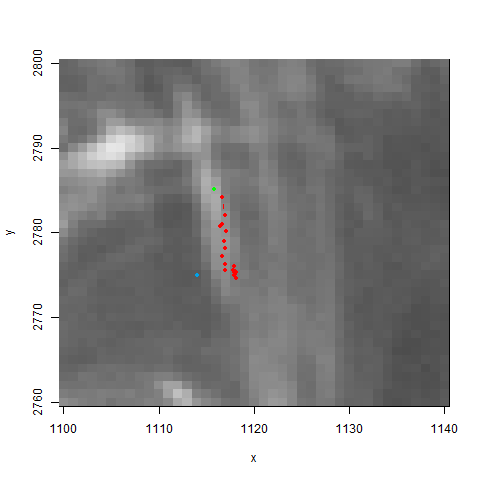}
  \caption{Illustration of a centroid moving from time step 15 (left) to time step 20 (center) to time step 25 (right). The blue point indicates the start point of the centroid, and the green point indicates the current centroid estimate.}
  \label{path}
\end{figure*}

In this paper, we propose a method to locate and track small-scale flow patterns in order to estimate their dynamic properties.
 The flow tracking algorithm is outlined in Subsect. 2.1,  the estimation of the local background temperature and sound speed is discussed in Subsect. 2.2, and in Subsect.~2.3 we perform simultaneous statistical inference on the flow velocities in order to classified them as subsonic, sonic, or supersonic. The results and corresponding discussion are presented in Sect.~3 and Sect.~4 consists of conclusive remarks.

\section{Methodology}
\subsection{Plasma flow identification}

In our analysis we consider the active region (AR) 11429, which was visible on the solar surface from 4 March 2012 to 15 March 2012.  We used the high-resolution Extreme Ultraviolet (EUV) monitor images taken by AIA onboard the SDO. In order to locate and track the flow patterns, we considered the image data from the 171$\AA$ spectral line. 
The AIA images, which have a 12 second time cadence, were analyzed during two intervals of time: (i) on 6 March 2012 from 12:22:00UT to 12:35:00UT when the active region was situated at -40 degrees from the central meridian of the visible solar disk (the limb data); and (ii) on 9 March 2012 from 03:00:00UT to 03:24:00UT when the active region was in the vicinity of the central meridian (the central data). On all AIA data we applied standard image preparation and cleaning techniques using the Solarsoftware (SSW) package.

It is an arduous task to locate and track plasma flows in real time by observing an 171$\AA$ spectral line consisting of thousands of pixel intensities that change over time. We propose a method that automatically locates and tracks any amount of candidate plasma flows simultaneously over a spectral line, and yields the coordinates and times of each flow. 
This methodology is based on two properties of the plasma flows: (i) A local signature of a plasma blob must appear (as an enhanced intensity pattern) and (ii) it must move. We define $I_k(x_i,y_i)$ as the intensity at the coordinates $(x_i,y_i)$ at time $t_k$. Since movement is indicated by a change in intensity at the specific coordinates $(x_i,y_i)$, we define the change in intensity as $\Delta I_k(x,y)=I_k(x,y)-I_{k-1}(x,y)$. The two properties of plasma flow mentioned above indicate that a combination of movement $\Delta I_k$ and intensity $I_k$ is considered when locating and tracking a flow. Therefore, we consider the linear combination $\gamma I_k(x,y)+(1-\gamma) \Delta I_k (x,y)$ for some $0 \leq \gamma \leq 1$ as the quantity used to locate and track plasma flows. For time steps $k=2,3,4,\dots$ we perform the following iterative procedures: 

\noindent  Step 1. Introducing new intensity enhancements and tracking candidate flows \\
We consider the M points (pixels) that have the largest values for $\gamma I_k(x,y)+(1-\gamma) \Delta I_k (x,y)$ and denote these as $(x_{1,k}, y_{1,k}), (x_{2,k}, y_{2,k}) \dots, (x_{M,k},y_{M,k})$. In order to introduce the candidate intensity enhancements, we execute the following iterative procedure for $i=1,2,\dots,M$: \\
Let $\mathbb{S}_{i,k} = \left\{ (x,y) : \sqrt{(x-x_{i,k})^2+(y-y_{i,k})^2} \leq 3\right\}$ and calculate
\begin{equation}
x_{i,k}^* = \frac{ \sum\limits_{(x,y)\in\mathbb{S}_{i,k}} x\, e^{\lambda \left( I_k(x,y) -\gamma \, I_{k-1}(x,y)\right) }}{\sum\limits_{(x,y)\in\mathbb{S}_k}e^{\lambda \left( I_k(x,y) -\gamma \, I_{k-1}(x,y)\right) }}
\end{equation}
\begin{equation}
y_{i,k}^* = \frac{\sum\limits_{(x,y)\in\mathbb{S}_k} y\, e^{\lambda \left( I_k(x,y) -\gamma \, I_{k-1}(x,y)\right) }}{\sum\limits_{(x,y)\in\mathbb{S}_k} e^{\lambda \left( I_k(x,y) -\gamma \, I_{k-1}(x,y)\right) }}  \label{iterative}
\end{equation}
If $\sqrt{(x_{i,k}^*-x_{i,k})^2+(y_{i,k}^*-y_{i,k})^2} > 0.01$ then set $(x_{i,k},y_{i,k}) = (x_{i,k}^*, y_{i,k}^*)$ and recalculate $\mathbb{S}_{i,k}, x_{i,k}^* $ and $y_{i,k}^* $.\\
Otherwise if $\sqrt{(x_{i,k}^*-x_{i,k})^2+(y_{i,k}^*-y_{i,k})^2} \leq 0.01,$ then the centroid has converged. \\

Further, to enable the tracking of the advance of these introduced enhancements, we perform the same iterative procedure as above for $i=M+1,M+2,\dots,M+N_{k-1}$ , where it is implied that $(x_{1,k-1},y_{1,k-1}), (x_{2,k-1},y_{2,k-1}), \dots (x_{N_{k-1},k-1}, y_{N_{k-1},k-1})$ denote the previous time's candidate flow points, where $N_1=0$ and $(x_{i+M,k}, y_{i+M,k}) = (x_{i,k-1},y_{i,k-1})$ for $i=1,2,\dots,N_k$ are set, where $N_k$ is the number of candidate flows. \\

The hyper-parameters $\gamma$ and $\lambda$ directly affect how the method locates and tracks flows. The parameter $\gamma$ specifies what feature is followed by the method, and the parameter $\lambda$ specifies how the method follows it.
For values of $\gamma$ close to 0, the method will only follow the largest changes in intensity, and for values close to 1, the method will only follow the largest intensity enhancements. A central compromise of $\gamma=\frac{1}{2}$ is recommended since this balances the requirement for the method to follow large intensities and also follow changes in these intensities. For values of $\lambda$ close to 0, the formula for $(x^*,y^*)$ will yield a point that is at the local mean of $ I_k(x,y) -\gamma \, I_{k-1}(x,y) $, and for a large value of $\lambda$, the formula for $(x^*,y^*)$ will give a point that is at the local maximum of $ I_k(x,y) -\gamma \, I_{k-1}(x,y) $. In general, if it is required that the maximum of $ I_k(x,y) -\gamma \, I_{k-1}(x,y) $ be given a weight $\omega$, then it follows that the parameter $\lambda$ be chosen as  $\lambda_\omega$  , which is the solution to the equation
\begin{equation}
\frac{\lambda_\omega \, \max\limits_{(x,y)\in\mathbb{S}_{i,k}} \left\{ I_k(x,y) -\gamma \, I_{k-1}(x,y) \right\} }{\log\sum\limits_{(x,y)\in\mathbb{S}_{i,k}} e^{\lambda_\omega \, \left( I_k(x,y) -\gamma \, I_{k-1}(x,y) \right) }} = \log \omega
.\end{equation}
It is recommended that $\lambda=\lambda_{\frac{1}{2}}$ be chosen so that half of the weight is given to the local maximum of $ I_k(x,y) -\gamma \, I_{k-1}(x,y) $. This, therefore, provides a compromise between a mean and maximum method of following a flow. The result of this iterative procedure, given by Eq.~(\ref{iterative}), is illustrated in Fig.~\ref{converge}. While the resulting movement of a candidate flow over time steps $k=2,3,4,\dots$ is illustrated in Fig.~\ref{path}.\\

\noindent  Step 2. The candidate flow assessment \\
The change in direction and change in distance of each of the candidate flows are assessed in order to determine whether they constitute a flow or not.
We denote the direction ($\theta_{i,k}$) and velocity ($v_{i,k})$ of each candidate flow $i=1,2,\dots M+N_{k-1}$ as 
\begin{equation}
\theta_{i,k} = \arctan \frac{y_{i,k}-y_{i,k-1}}{x_{i,k}-x_{i,k-1}}
\end{equation}
\begin{equation}
v_{i,t} = \frac{\sqrt{(x_{i,k}-x_{i,k-1})^2+(y_{i,k}-y_{i,k-1})^2}}{t_{i,k}-t_{i,k-1}}.
\label{assess}
\end{equation}
Consequently, a candidate flow is disregarded and removed if $|\Delta\theta_{i,k}|>\Theta$ or $v_{i,k} \notin [v_{min}, v_{max}]$. \\

Altogether this method has four dummy parameters ($M, v_{min}, v_{max}$ , and $\Theta$) and two hyper-parameters ($\gamma$ and $\lambda$) that are set by the user and they output a set of flow position coordinates $\left\{(x_{i,k},y_{i,k}) \right\}_{k=k_i, k_i+1, \dots, k_i+K_i }$ and occurrence times $\left\{t_{i,k}\right\}_{k=k_i, k_i+1, \dots, k_i+K_i }$ for $i=1,2,\dots,N$, where $N$ is the total number of identified flows, $K_i$ is the number of time steps of the $i^{th}$ flow and $k_i$ is the start time step of the $i^{th}$ flow. Ideally, $M$ would be chosen to be as large as possible so that the greatest amount of candidate flows can be identified. The velocity parameters ($v_{min}$ and $v_{max}$) control the minimum and maximum velocity of the candidate flows, and should be set to identify  flows with reasonable velocities. Alternatively, if the user is interested in flows of a specific velocity, then these parameters can be initialized accordingly. The direction parameter $\Theta$ controls the maximum change in direction that a candidate flow is allowed to have at each observed time. The ranges of the minimum and maximum velocity and the angle $\Theta$ approve the candidate flows and filter out all others that do not satisfy the required conditions. \\

\begin{figure*}[h!]
  \includegraphics[width=177mm, trim={5mm 0 5mm 0}, clip]{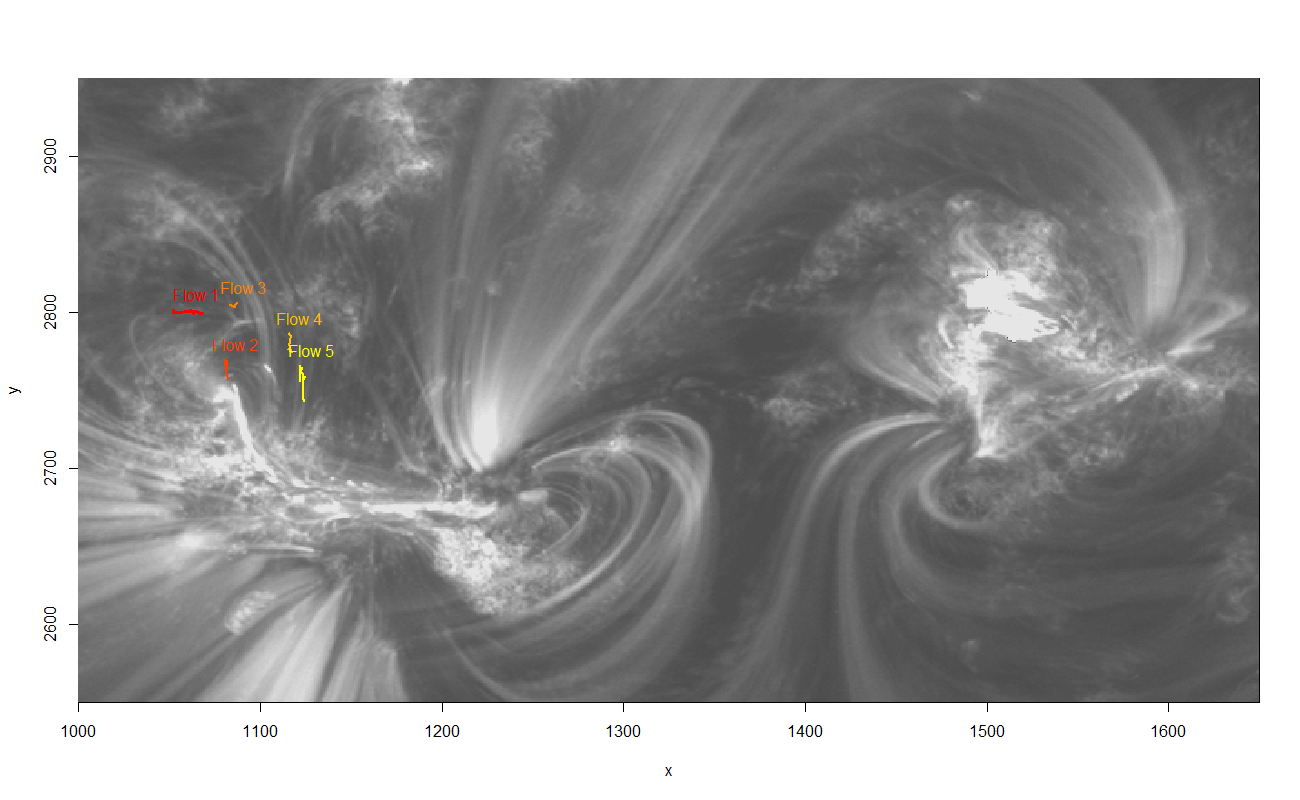}
    \includegraphics[width=177mm, trim={5mm 0 5mm 0}, clip]{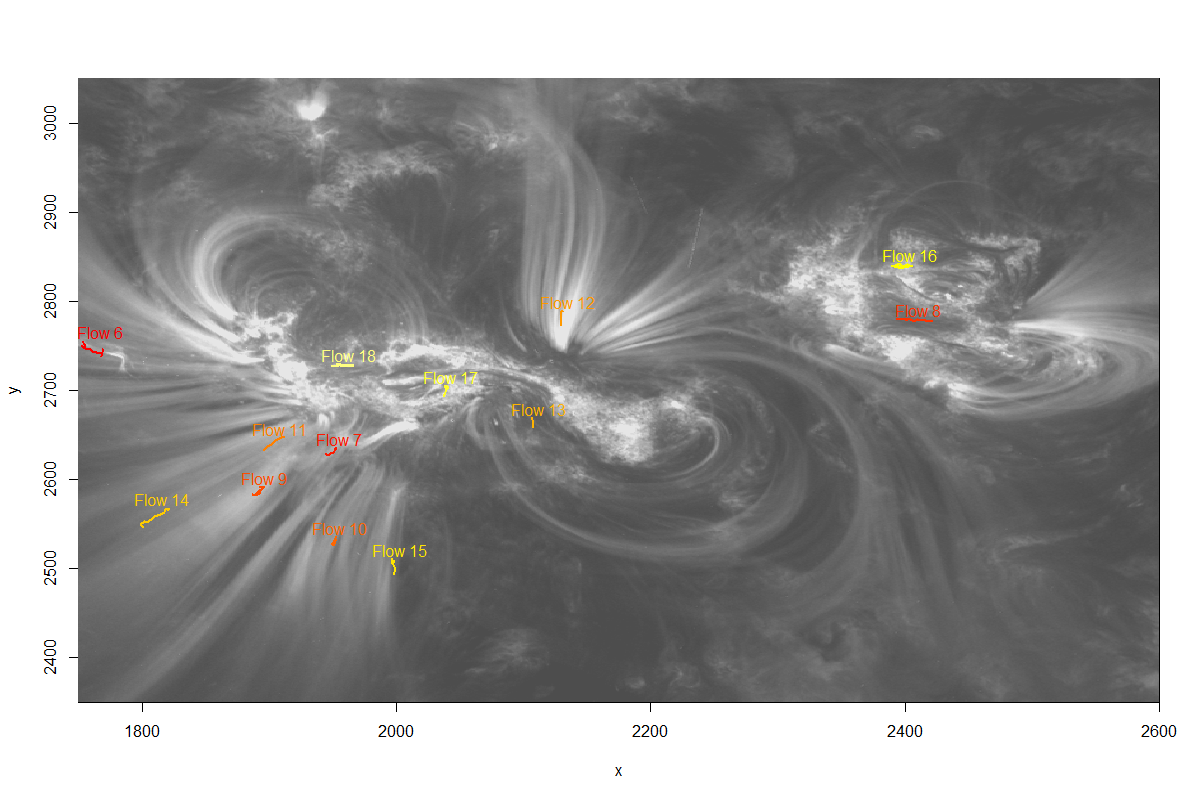}
  \caption{Illustration of the flows identified by applying the proposed methodology to the SDO image data from the 171$\AA$ spectral line for the date 6 March 2012 from 12:22:00 to 12:35:00 referred to as the limb data (top) and again for the date 9 March 2012 from 03:00:00 to 03:24:00 referred to as the central data (bottom).}
  \label{global}
\end{figure*}

The proposed methodology is applied to the SDO image data from the 171$\AA$ spectral line for the limb data and again for the central data. The dummy parameters are chosen as $M=100$,  $v_{min}=40 km/s$, $v_{max}= 200 km/s,$ and $\Theta=\pi/6$. The hyper-parameters are set as $\gamma=0.5$ and $\lambda=\lambda_{\frac{1}{2}}$ as we proposed above. The application of the methodology yielded a total of 5 flows for the limb data and a further 13 flows for the central data. These results are illustrated in Fig.~\ref{global} and are summarized in Table~\ref{flowtable}. \\

The method can easily be extended to locate and track flows over multiple spectral lines simultaneously by substituting  $\lambda \left( I_k(x,y) -\gamma \, I_{k-1}(x,y)\right)$ with $\sum_f \lambda_{f} \left( I^{f}_k(x,y) -\gamma \, I_{k-1}^f(x,y)\right)$ in the iterative procedure, where $f$ indicates the spectral lines of interest. This would require the specification of an additional parameter value $\lambda_f$ for each additional spectral line $f$.

\subsection{DEM analysis and background sound speed estimation}
We estimate the mean background temperature for each flow by applying the DEM inversion method  of \cite{Hannah2012}
to the six AIA channels, fitting Gaussian functions to the resulting DEM curves and taking the mean of the peaks of the fitted Gaussian functions.
Following the algorithm developed by \citet{Plowman2013}, for each flow we take the vector of pixels (the set of pixels are selected by the iterative procedure described above) along its trajectory at the time preceding the appearance of the intensity enhancement (blob) and extract the corresponding six AIA channel values along it. These six AIA channels are 94$\AA$, 171$\AA$, 193$\AA$, 211$\AA$, 304$\AA,$  and 335$\AA$. We then apply the DEM inversion method  to obtain reconstructed DEM distributions for each flow, respectively. 
Gaussian functions are fitted to the DEM curves to estimate the peak temperature for each pixel at each time step. 
We then take the mean of these peak values of the temperature for the entire time span from the start of the observation until the occurrence of the flow to estimate the local background temperature.
The local background sound speed of each flow can be estimated by using the  formula
\begin{equation}
\centering
Cs={\sqrt{\frac{\gamma k_{B} T}{\tilde{\mu} m_{p}}}},
\label{Cs}
\end{equation}
where $\mu=0.6$ is the mean atomic mass factor of hydrogen plasma, $\gamma=5/3$, $k_{B}$ is the Bolzman constant, and $T$ is the estimated background temperature.
The error of the estimated background temperature $T$ can be estimated by the shape parameter $\sigma_T$ of the Gaussian functions fitted to the DEM curves \citep{Plowman2013}. Equation \ref{Cs} is again used to estimate the error of the resulting background sound speed, denoted by $\hat{\sigma}_{C_s}$. The estimated background sound speeds and their corresponding errors are summarized in Table \ref{flowtable}.

\subsection{Flow velocity inference}
In this section, we describe the statistical inference of the flow velocity detection and their classification. However, before doing this, it is reasonable to note that the 2D AIA images  comply with a deformation of the spherical surface image of the Sun due to the projection effect. In order to take into account this deformation and recalculate the distances covered by AIA pixels, we use the approach that each pixel projection on the solar surface along the line of sight represents a locally planar non-rectangular section of the solar surface. If one then performs straightforward transformations of coordinates, the solution of this geometrical ansatz (derivation of it is rather cumbersome but straightforward, therefore for the sake of brevity we just demonstrate here the final result), it can be shown that the mentioned non-rectangular area would be seen as a parallelogram if it were situated right in the center of the visible solar disk image with the diagonals
\begin{equation}
M_1 = \frac{L}{ \cos(\theta)}\sqrt{2+\left(\tan(\theta)+\frac{\tan(\phi)}{\cos(\theta)}\right)^2}
\end{equation}
\begin{equation}
M_2 = \frac{L}{ \cos(\theta)}\sqrt{2+\left(\tan(\theta)-\frac{\tan(\phi)}{\cos(\theta)}\right)^2}.
\end{equation}
The angle between them is defined as
\begin{equation}
\cos(\alpha)  = \frac{M_1^2 + M_2^2 - 4L^2 }{ 2M_1M_2},
\end{equation}
where, $L=R_0/R$ is the standard length of the pixel side, which is determined by the ratio of solar radius $R_0$ in kilometers  and the actual solar radius $R$ measured in pixels seen by the observer at the current moment in time. The angles $\phi$ and $\theta$ represent, respectively, the angular longitudinal and latitudinal distance of the pixel from the center of the solar disk, also with respect to\ the observer's position, which are defined as
\begin{equation}
\sin(\phi)= \frac{x-CRPIX1}{R},\hspace{15mm}
\sin(\theta)= \frac{y-CRPIX2}{R}.
\end{equation}
Here, $x$ and $y$ are pixel coordinates as defined above and $CRPIX1$ and $CRPIX2$ are the corresponding pixel coordinates of the visible solar center on the $AIA\_PREP$-ed SDO image. From these definitions, the sides of the obtained parallelogram can be defined as
\begin{equation}
L_1 ^{\prime}=\sqrt{\frac{M_1^2+M_2^2}{4}-\frac{M_1M_2}{2}\cos(\alpha)}
\end{equation}
\begin{equation}
L_2 ^{\prime}=\sqrt{\frac{M_1^2+M_2^2}{4}+\frac{M_1M_2}{2}\cos(\alpha)}.
\end{equation}
We take the smallest side of the parallelogram $L_{min}=L_1 ^{\prime}$ and the largest diagonal $L_{max}=\max\{M_1,M_2\}$ as the minimum and maximum length scales, which are used for the estimation of the lower and upper bounds of the measured velocities. In this study, we do not perform any 3D reconstruction of the magnetic field. Therefore, another projection effect occurs related to the angle between the local orientation of the loop sustaining the observed flow and the line of sight. Although, the additional effect caused by the orientation of the magnetic field only can lead to the conclusion that the actual velocities  can never be smaller (but they are equal or somewhat larger) than either the lower and upper bounds of them. The latter are estimated throughout the current work. \\

The above specifications imply that the observed velocities $v_1, v_2, \dots, v_k$ are bounded by minima $v_1^{min}, v_2^{min},\dots,v_k^{min}$ and maxima $v_1^{max}, v_2^{max},\dots,v_k^{max}$ where
\begin{equation}
v_{j}^{min} =\frac{L_{min}\sqrt{(x_{j}-x_{j-1})^2+(y_{j}-y_{j-1})^2}}{t_{j}-t_{j-1}}
\end{equation}
\begin{equation}
v_{j}^{max} = \frac{L_{max}\sqrt{(x_{j}-x_{j-1})^2+(y_{j}-y_{j-1})^2}}{t_{j}-t_{j-1}},
\end{equation}
where $j=1,2,...,k$.\\

In order to test whether a the apparent velocity of a flow is sonic, subsonic, or supersonic at time $t$ , we perform the following hypothesis tests simultaneously:
\begin{equation}
H_{01}: v^{max}(t) = C_s \hspace{3mm} H_{11}: v^{max}(t) < C_s \hspace{3mm} H_{21}: v^{max}(t) > C_s 
\end{equation}
\begin{equation}
H_{02}: v^{min}(t) = C_s \hspace{3mm} H_{12}: v^{min}(t) > C_s \hspace{3mm} H_{22}: v^{min}(t) < C_s
,\end{equation}
 where $v^{max}(t)$ and $v^{min}(t)$ are the upper and lower flow velocity bounds at time $t$ and $C_s$ is its background sound speed. The flow velocity upper and lower bounds $v^{max}$ and $v^{min}$ are approximated by a quadratic regression model function of time,
\begin{equation}
v^{max}(t) = \beta_0^{max} + \beta_1^{max} \, t + \beta_2^{max} \, t^2 + \epsilon^{max},
\end{equation}
\begin{equation}
v^{min}(t) = \beta_0^{min} + \beta_1^{min} \, t + \beta_2^{min} \, t^2 + \epsilon^{min},
\end{equation}
where $\beta_0^{max}$, $\beta_1^{max}$, $\beta_2^{max}$, $\beta_0^{min}$, $\beta_1^{min}$ , and $\beta_2^{min}$  are real-valued coefficients and $\epsilon^{max}$ and $\epsilon^{min}$ are the error terms that are assumed to be normally distributed with mean $0$ and standard deviation $\sigma_{\epsilon^{max}}$ and $\sigma_{\epsilon^{min}}$ respectively. The least squares estimators of $v^{max}(t)$ and $v^{min}(t)$ are given by \citep{Hastie}
\begin{equation}
\hat{v}^{max}(t)
        = \left[ \begin{array}{c}1 \\ t \\ t^2 \end{array} \right]^T \, \left( \left[ \begin{array}{c}\underline{1} \\ \underline{t} \\ \underline{t}^2 \end{array} \right]^T\, \left[ \begin{array}{c}\underline{1} \\ \underline{t} \\ \underline{t}^2 \end{array} \right]\right)^{-1} \, \left[ \begin{array}{c}\underline{1} \\ \underline{t} \\ \underline{t}^2 \end{array} \right]^{T} \, \underline{v}^{max}, 
        \label{leastsquaresmax}
\end{equation}
\begin{equation}
\hat{v}^{min}(t) 
        = \left[ \begin{array}{c}1 \\ t \\ t^2 \end{array} \right]^T \, \left( \left[ \begin{array}{c}\underline{1} \\ \underline{t} \\ \underline{t}^2 \end{array} \right]^T\, \left[ \begin{array}{c}\underline{1} \\ \underline{t} \\ \underline{t}^2 \end{array} \right]\right)^{-1} \, \left[ \begin{array}{c}\underline{1} \\ \underline{t} \\ \underline{t}^2 \end{array} \right]^{T} \, \underline{v}^{min}, 
        \label{leastsquaresmax}
\end{equation}
where $\underline{t} = [t_1, t_2, \dots, t_K]^T$, $\underline{t}^2 = [t_1^2, t_2^2, \dots, t_K^2]^T$, $\underline{v}^{max}=[v_1^{max}, v_2^{max}, \dots v_K^{max}]^T$ and $\underline{v}^{min}=[v_1^{min}, v_2^{min}, \dots v_K^{min}]^T$ are the vectors of observational moments of time, the squares of them, and the corresponding observed values of the velocity upper and lower bounds respectively.

\begin{table*}[h!]
\caption{Overview of the identified flows together with their respective duration, sound speed, and the standard errors of the estimated fitted upper and lower velocity bounds.}
\label{flowtable}
\centering
\begin{tabular}{l | c c c c c c c c}
\hline \hline
Flow    & Date          & Start time & End time & Duration (s) & $C_s (km/s) $       & $\hat{\sigma}_{C_s}  (km/s)$  &   $\hat{\sigma}_{\epsilon^{max}} (km/s)$         & $\hat{\sigma}_{\epsilon^{min}} (km/s)$ \\
\hline 
1               & 6/3/2012      & 12:24:24      & 12:25:12      &  48 & 90.34 &7.12 &                 4.38 & 9.32 \\ 
2               & 6/3/2012      & 12:23:48      & 12:24:36      & 48 & 90.09 & 6.26 &                        24.90 & 55.53 \\
3               & 6/3/2012      & 12:26:48      & 12:27:36      & 48 & 90.08 & 6.98 &                        1.19 & 2.67\\
4               & 6/3/2012      & 12:25:48      & 12:26:36      & 48 & 89.96 & 8.17 &                        15.86 & 32.64 \\
5               & 6/3/2012      & 12:25:12      & 12:27:12      & 120 & 90.07 & 9.22 &                31.08 & 63.72\\
6               & 9/3/2012      & 03:03:34      & 03:04:34      & 60 & 90.87 & 7.44 &        25.76 & 38.48 \\
7               & 9/3/2012      & 03:02:34      & 03:03:34      & 60 & 90.64 & 8.83 &        25.23 & 40.14\\         
8               & 9/3/2012      & 03:05:36      & 03:06:36      & 60 & 91.34 &  9.01 &       23.09 & 37.60 \\
9               & 9/3/2012      & 03:07:12      & 03:08:24      & 72 & 91.74 & 8.86 &                        29.85 & 44.65\\
10              & 9/3/2012      & 03:07:24      & 03:08:24      & 60 & 90.88 &  7.02 &                       8.88 & 13.05 \\
11              & 9/3/2012      & 03:12:12      & 03:13:12      & 60 & 91.03 & 8.81 &                        42.75 & 64.47\\
12              & 9/3/2012      & 03:14:24      & 03:15:36      & 72 & 91.01 & 9.40 &                        26.36 & 40.21\\
13              & 9/3/2012      & 03:15:00      & 03:16:12      & 72 & 90.17 & 5.61 &                         13.57 & 20.14\\
14              & 9/3/2012      & 03:19:48      & 03:20:48      & 60 & 91.35 & 7.93 &                         6.24 & 9.41\\
15              & 9/3/2012      & 03:21:00      & 03:22:00      & 60 & 91.23 &  9.51 &                       15.56 & 22.65 \\
16              & 9/3/2012      & 03:20:36      & 03:21:36      & 60 & 90.93 &  6.36 &                       23.13 & 38.42\\
17              & 9/3/2012      & 03:21:00      & 03:22:00      & 60 & 91.31 &  9.25 &               25.79 & 38.41\\
18              & 9/3/2012      & 03:21:36      & 03:22:36      & 60 & 90.97 &       8.55 &          30.65 & 50.92\\
\hline 
\end{tabular}
\end{table*} 
 
Under $H_{01}$ we have that $\hat{v}^{max}(t) - C_s \sim N\left( 0 , \sigma_{C_s}^2 + \sigma^2_{\epsilon^{max}} \, \xi(t) \right)$ and under $H_{02}$ we have that $\hat{v}^{min}(t) - C_s \sim N\left( 0 ,\sigma_{C_s}^2+ \sigma^2_{\epsilon^{min}} \, \xi(t) \right)$ , where in both cases
\begin{equation}
\xi(t)=  \left[ \begin{array}{c}1 \\ t \\ t^2 \end{array} \right] \, \left( \left[ \begin{array}{c}\underline{1} \\ \underline{t} \\ \underline{t}^2 \end{array} \right]^T\, \left[ \begin{array}{c}\underline{1} \\ \underline{t} \\ \underline{t}^2 \end{array} \right]\right)^{-1} \,   \left[ \begin{array}{c}1 \\ t \\ t^2 \end{array} \right]^T.
\end{equation}
It follows that the test statistics under $H_{01}$ is given by
\begin{equation}
T^{max}(t) = \frac{\hat{v}^{max}(t) - C_s}{\sqrt{\hat{\sigma}_{C_s}^2+\hat{\sigma}_{\epsilon^{max}}^2 \xi(t)}} \sim t_{K-1}
,\end{equation}
and the test statistics under $H_{02}$ is given by
\begin{equation}
T^{min}(t) = \frac{\hat{v}^{min}(t) - C_s}{\sqrt{\hat{\sigma}_{C_s}^2+\hat{\sigma}_{\epsilon^{min}}^2 \xi(t)}} \sim t_{K-1}
,\end{equation}
where $t_{K-1}$ denotes the Student-t distribution with $K-1$ degrees of freedom and $\hat{\sigma}_{\epsilon^{max}}$     and $\hat{\sigma}_{\epsilon^{min}}$ denote the standard errors of $\hat{v}^{max}(t)$ and $\hat{v}^{min}(t)$ respectively, which are given by
\begin{equation}
\hat{\sigma}_{\epsilon^{max}} =\sqrt{\frac{1}{K} \, \sum_{k=1}^K \left(\hat{v}^{max}(t_k) - v_k^{max} \right)^2}  
\end{equation}
\begin{equation}
\hat{\sigma}_{\epsilon^{min}} = \sqrt{\frac{1}{K} \, \sum_{k=1}^K \left(\hat{v}^{min}(t_k) - v_k^{min} \right)^2} 
.\end{equation}

At time $t$ with a $100(1-\alpha)\%$ level of significance, $H_{02}$ is rejected and $H_{12}$ is accepted when $T^{min}(t) > t_{K-1}(1-\alpha)$), $H_{01}$ is rejected and $H_{11}$ is accepted when $T^{max}(t) < t_{K-1}(\alpha)$, $H_{01}$ is rejected and $H_{21}$ is accepted when $T^{min}(t) < t_{K-1}(\alpha),$ and $H_{02}$ is rejected and $H_{22}$ is accepted when $T^{max}(t) > t_{K-1}(1-\alpha)$). The amount $t_{K-1}(\alpha)$ denotes the $\alpha^{th}$ quantile of the Student-t distribution with $K-1$ degrees of freedom. 
These hypothesis tests form the basis on which a flow is classified as supersonic, subsonic, or sonic. 

\subsection{Classification of flows}
At time $t$ with a $100(1-\alpha)\%$ level of significance, a flow is classified as supersonic if  $H_{02}$ is rejected and $H_{12}$ is accepted, otherwise as subsonic if $H_{01}$ is rejected and $H_{11}$ is accepted, otherwise as sonic if $H_{01}$ is rejected and $H_{21}$ and $H_{02}$ is rejected and $H_{22}$ is accepted. At a $100(1-\alpha)\%$ level of significance, a flow can have different classifications at different times, and at other times no classification at all.
In order to determine the predominant classification of the flow at a $100(1-\alpha)\%$ level of significance, it is necessary to determine the cumulative durations that the flow was supersonic, sonic, subsonic, or none at this level of significance. 
Let $\Pi^{supersonic}(\alpha)$, $\Pi^{sonic}(\alpha),$ and $\Pi^{subsonic}(\alpha)$ be the cumulative durations that the flow is classified as supersonic, sonic, and subsonic respectively at a $100(1-\alpha)\%$ level of significance:
\begin{equation}
\Pi^{supersonic}(\alpha) = \int I \left\{T^{min}(t) > t_{K-1}(1-\alpha) \right\} \, dt 
\end{equation}
\begin{equation}
\Pi^{sonic}(\alpha) =  \int I \left\{ T^{min}(t) < t_{K-1}(\alpha) \right\} \, I \left\{T^{max}(t) > t_{K-1}(1-\alpha) \right\} \, dt 
\end{equation}
\begin{equation}
\Pi^{subsonic}(\alpha) = \int I \left\{ T^{max}(t) < t_{K-1}(\alpha) \right\} \, dt 
,\end{equation}
where operator $I$ denotes indicator function. 

A flow is said to be predominantly supersonic, sonic, or subsonic at a $100(1-\alpha)\%$ level of significance if, respectively, $\Pi^{supersonic}(\alpha)$, $\Pi^{sonic}(\alpha),$ or $\Pi^{subsonic}(\alpha)$ is the largest. However, if both $\Pi^{supersonic}(\alpha)$ and $\Pi^{subsonic}(\alpha)$ are of a similar size then it is not clear which classification is predominant. In such a case, the flow is said to be transonic since it transitions from one predominant classification to another over time.

The predominant classification of a flow depends crucially on two properties of the estimated velocity bounds: their closeness to the background sound speed $C_s$ and the size of their respective standard errors $\hat{\sigma}_{\epsilon^{max}}$ and $\hat{\sigma}_{\epsilon^{min}}$. If $\hat{v}^{min}(t)-C_s$ is larger then $T^{min}(t)$ is larger, which means that the flow is classified as supersonic at a higher level of significance at time $t$.  If $C_s - \hat{v}^{max}(t)$ is larger then $-T^{max}(t)$ is larger, which means that the flow is classified as subsonic at a higher level of significance at time $t$. If both $C_s-\hat{v}^{min}(t)$ and $\hat{v}^{max}(t)-C_s$ are larger then both $-T^{min}(t)$ and $T^{max}(t)$ are larger, which means that the flow is classified as sonic at a higher level of significance at time $t$. The size of the standard errors $\hat{\sigma}_{\epsilon^{max}}$ and $\hat{\sigma}_{\epsilon^{min}}$ inversely increase the level of significance of the flow's classification. If the standard errors are larger, then the flow will be classified at a lower level of significance, and vice versa.

\begin{table*}[ht!]
\caption{Overview of the identified flows and their predominant classification at a $70\%$ level of significance}
\label{flowtable70}
\centering
\begin{tabular}{l | c c c c}
\hline \hline
Flow    & $\Pi^{supersonic}(0.7) $  &  $\Pi^{sonic}(0.7) $      &  $\Pi^{subsonic}(0.7) $       & Classification \\
\hline 
1 & 19 & 10.4 & 0.6 & Supersonic \\ 
2 & 7 & 0 & 4.2 & Supersonic * \\ 
3 & 0 & 0 & 34.6 & Subsonic \\ 
4 & 0 & 5.4 & 4.6 & Sonic \\ 
5 & 49 & 22 & 0 & Supersonic \\ 
6 & 0 & 0 & 25 & Subsonic \\ 
7 & 15.4 & 1 & 1.8 & Supersonic \\ 
8 & 0 & 8 & 0 & Sonic \\ 
9 & 5 & 1.6 & 0 & Supersonic \\ 
10 & 10.6 & 3.8 & 29 & Subsonic * \\ 
11 & 44.8 & 0 & 0 & Supersonic \\ 
12 & 33.2 & 0 & 0 & Supersonic \\ 
13 & 0 & 19.8 & 10.4 & Sonic \\ 
14 & 48.2 & 0 & 0 & Supersonic \\ 
15 & 31 & 2.6 & 0 & Supersonic \\ 
16 & 32.2 & 0 & 0 & Supersonic \\ 
18 & 24.2 & 0 & 0 & Supersonic\\
\hline
\end{tabular}
\end{table*} 

\begin{table*}[ht!]
\caption{Overview of the identified flows and their predominant classification at a $80\%$ level of significance}
\label{flowtable80}
\centering
\begin{tabular}{l | c c c c}
\hline \hline
Flow    & $\Pi^{supersonic}(0.8) $  &  $\Pi^{sonic}(0.8) $      &  $\Pi^{subsonic}(0.8) $       & Classification \\
\hline 
1 & 16.8 & 8.2 & 0 & Supersonic \\ 
2 & 4.6 & 0 & 0 & Supersonic \\ 
3 & 0 & 0 & 34 & Subsonic \\ 
5 & 39 & 9.4 & 0 & Supersonic \\ 
6 & 0 & 0 & 11.6 & Subsonic \\ 
10 & 10 & 2 & 27.8 & Subsonic * \\ 
11 & 42 & 0 & 0 & Supersonic \\ 
12 & 18.6 & 0 & 0 & Supersonic \\ 
13 & 0 & 5 & 8.4 & Subsonic \\ 
14 & 48.2 & 0 & 0 & Supersonic \\ 
15 & 24 & 0 & 0 & Supersonic \\ 
16 & 23.2 & 0 & 0 & Supersonic \\ 
\hline
\end{tabular}
\end{table*} 

\begin{table*}[ht!]
\caption{Overview of the identified flows and their predominant classification at a $90\%$ level of significance}
\label{flowtable90}
\centering
\begin{tabular}{l | c c c c}
\hline \hline
Flow    & $\Pi^{supersonic}(0.9) $  &  $\Pi^{sonic}(0.9) $      &  $\Pi^{subsonic}(0.9) $       & Classification \\
\hline 
1 & 7 & 2.8 & 0 & Supersonic \\ 
3 & 0 & 0 & 30.8 & Subsonic \\ 
10 & 8.8 & 0 & 23.6 & Subsonic * \\ 
11 & 29.2 & 0 & 0 & Supersonic \\ 
13 & 0 & 0 & 3.8 & Subsonic \\ 
14 & 47.8 & 0 & 0 & Supersonic \\ 
\hline
\end{tabular}
\end{table*} 

\section{Results and discussion}
\begin{figure*}[h!]
\includegraphics[width=6cm]{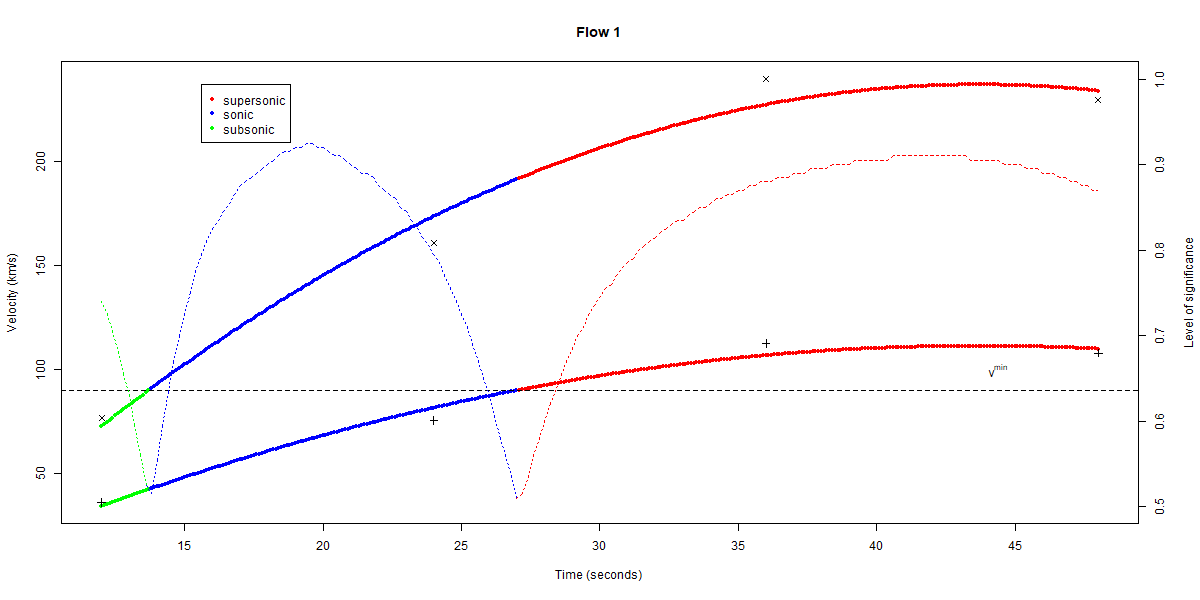}
\includegraphics[width=6cm]{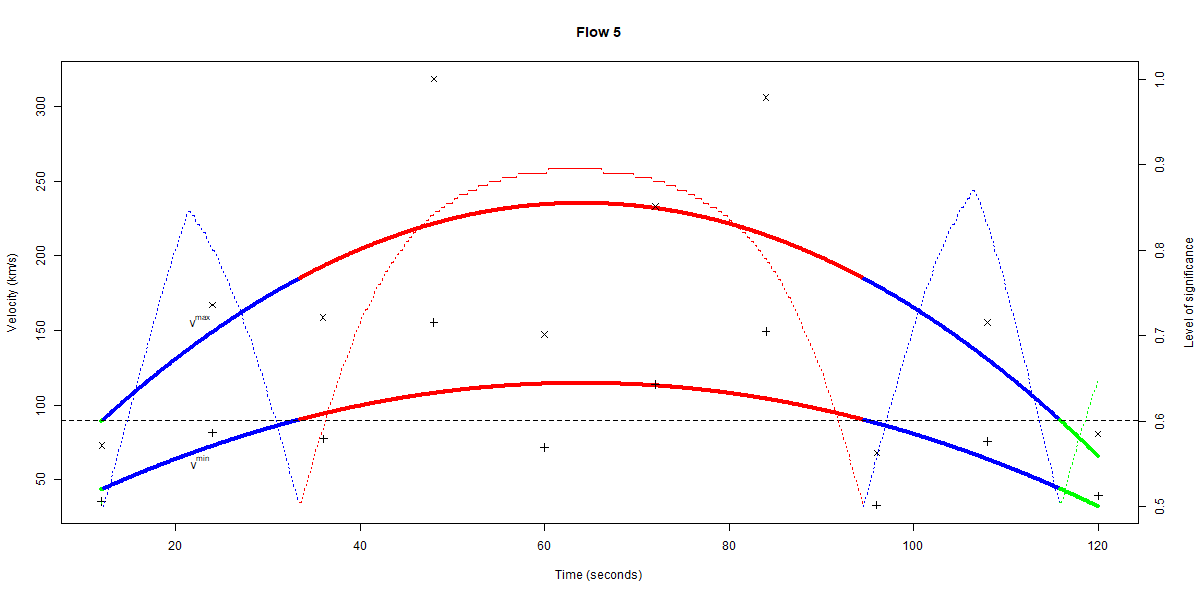}
        \includegraphics[width=6cm]{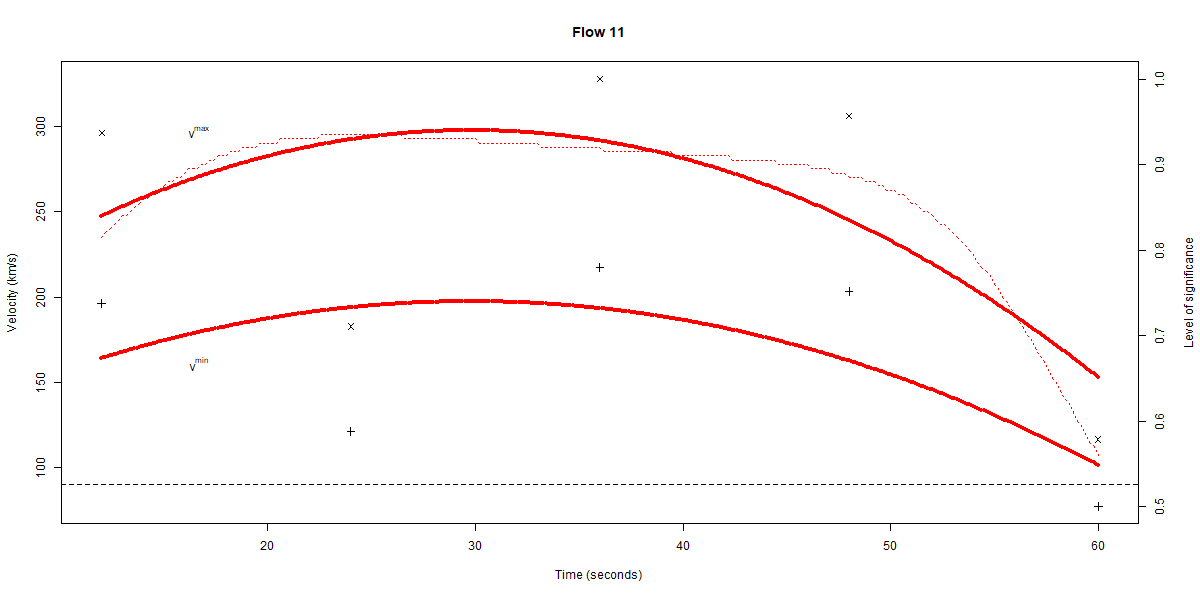}
        \includegraphics[width=6cm]{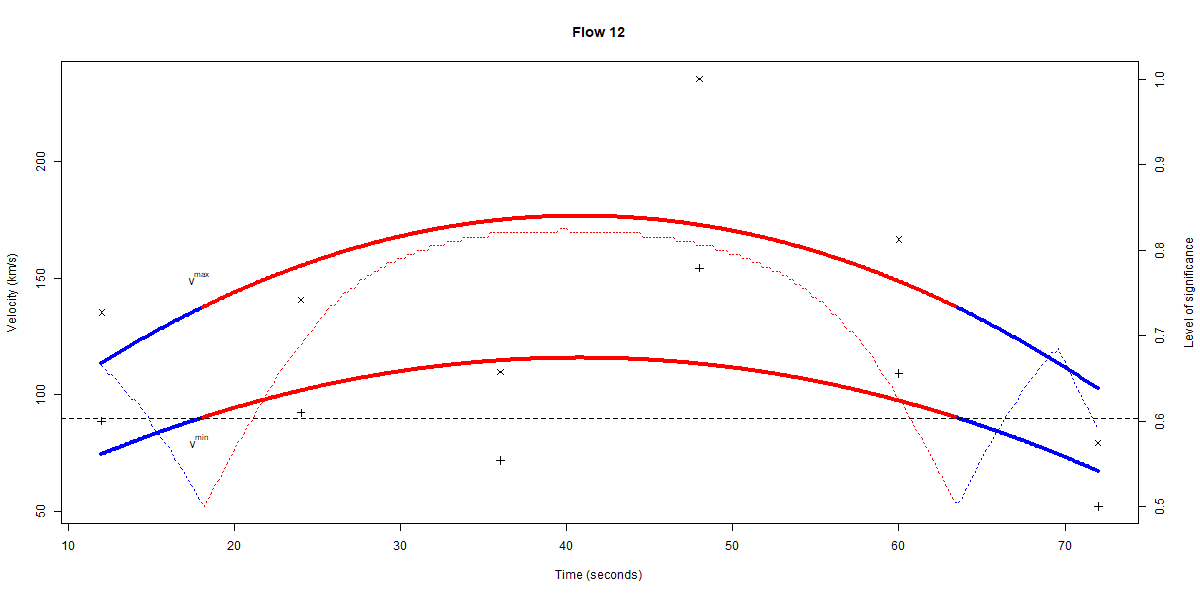}
        \includegraphics[width=6cm]{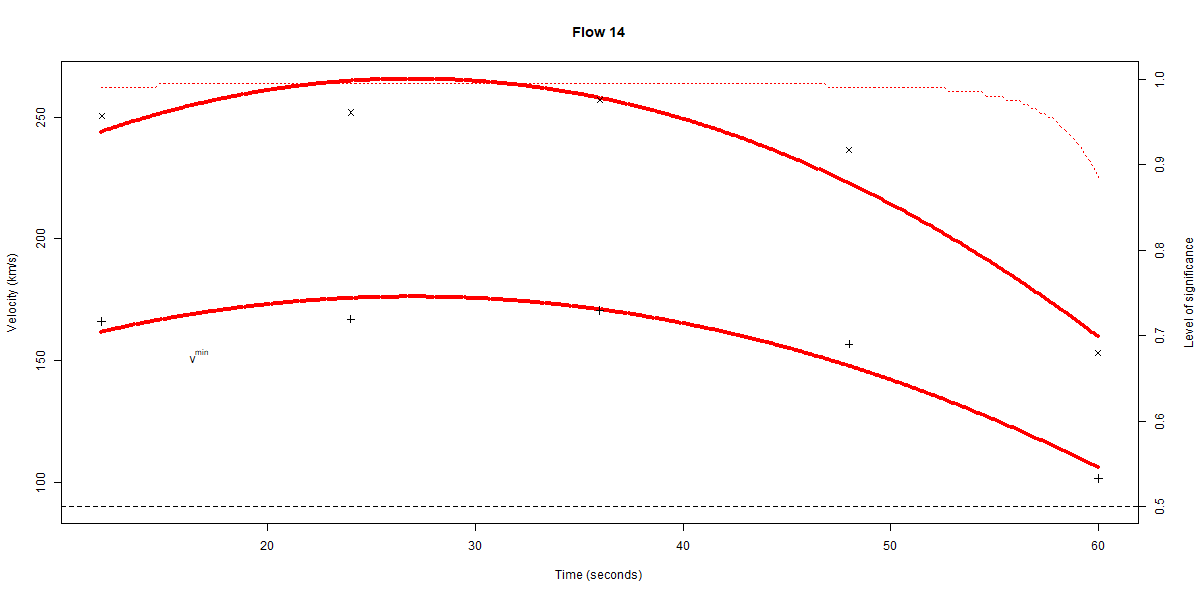}
        \includegraphics[width=6cm]{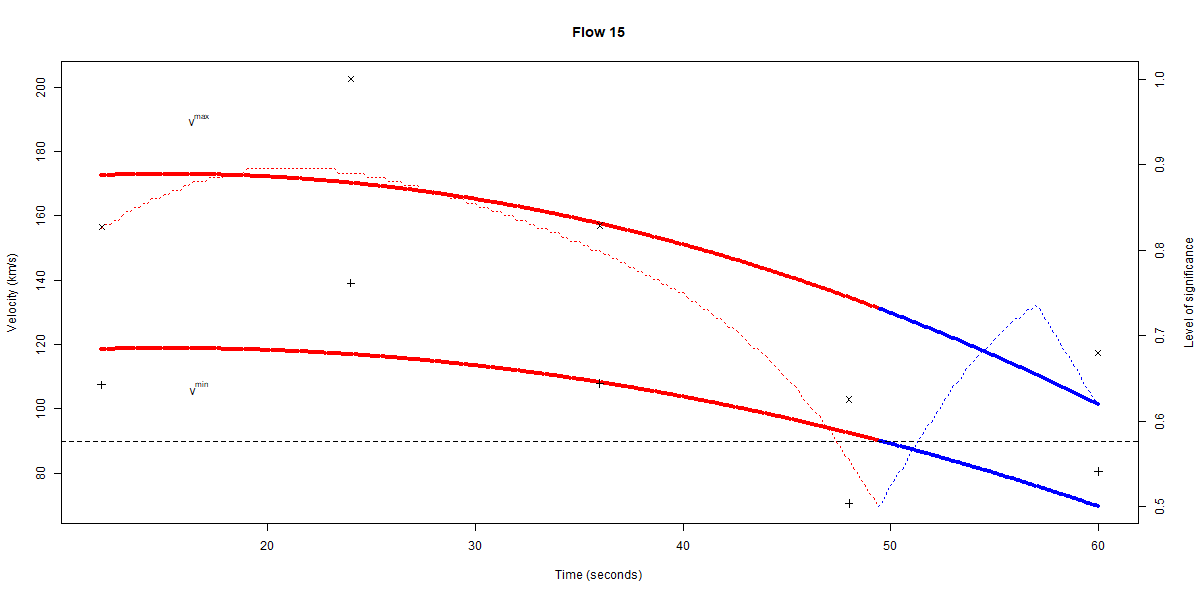}
        \includegraphics[width=6cm]{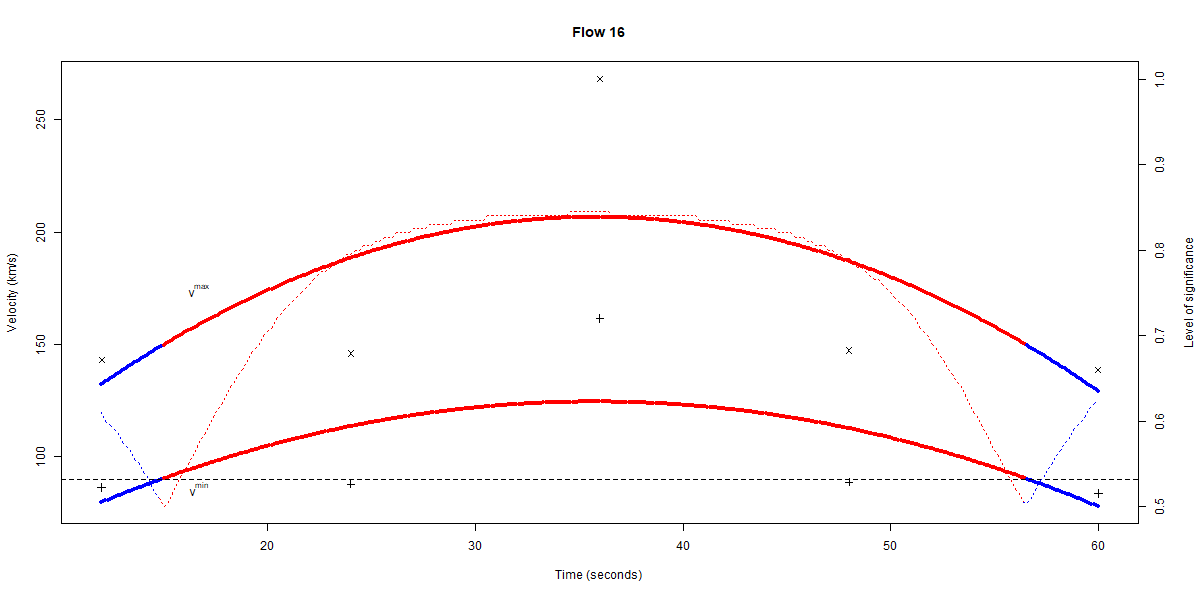}
        \includegraphics[width=6cm]{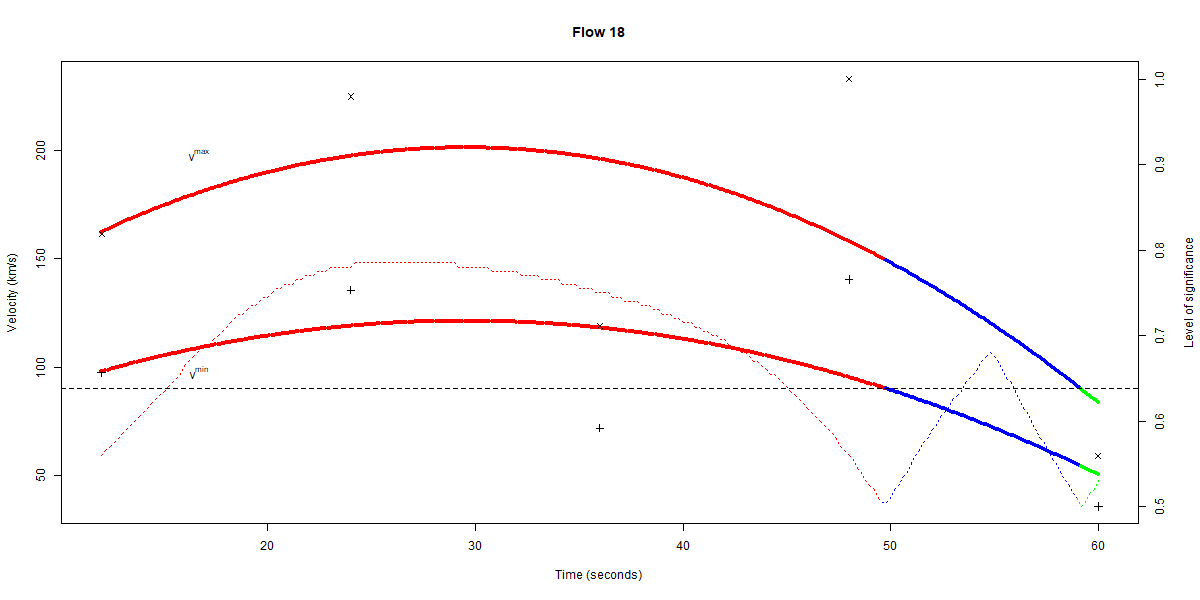}
  \caption{Velocity lower bound $\hat{v}^{min}$ and upper bound $\hat{v}^{max}$ together with the classification level of significance $100(1-\alpha)\%$. The red, blue, and green (solid line) curves indicate a supersonic, sonic, and subsonic classification (left ordinate axes), respectively  at a $50\%$ level of significance. The dotted line curves show the level of significance (rights ordinate axes). The horizontal dashed line represents the background sound speed. These flows all have a majority supersonic classification.
   }
  \label{supersonic}
\end{figure*}

\begin{figure}[h!]
\includegraphics[width=6cm]{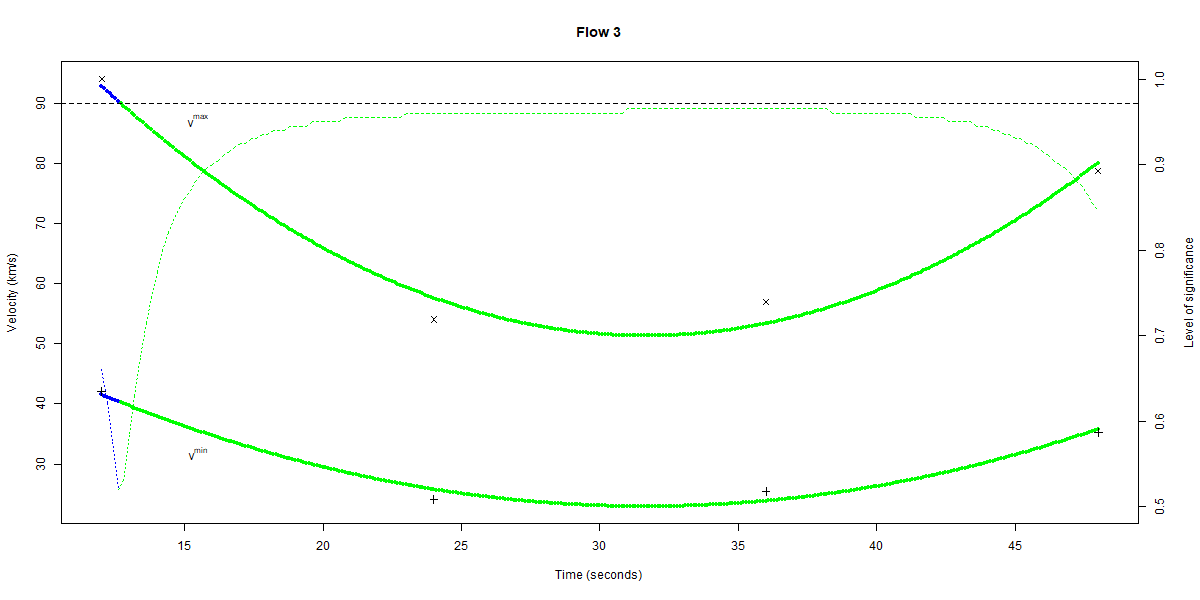}
\includegraphics[width=6cm]{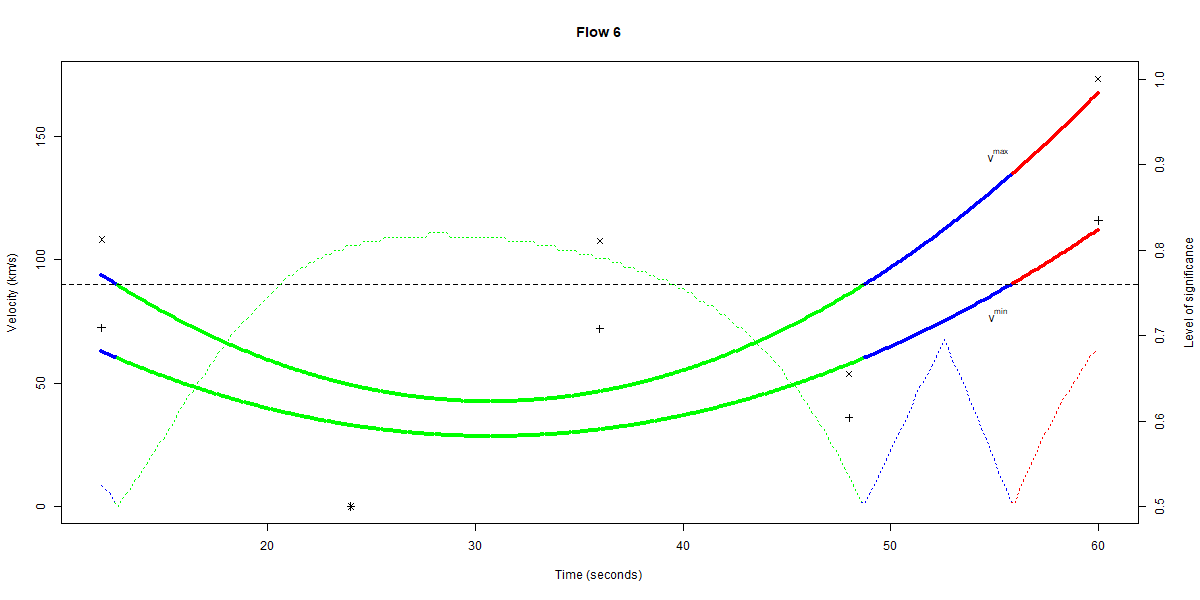}
\caption{Same as in Fig.~\ref{supersonic} for subsonic flow patterns.}
\label{subsonic}
\end{figure}

\begin{figure}[h!]
    \includegraphics[width=6cm]{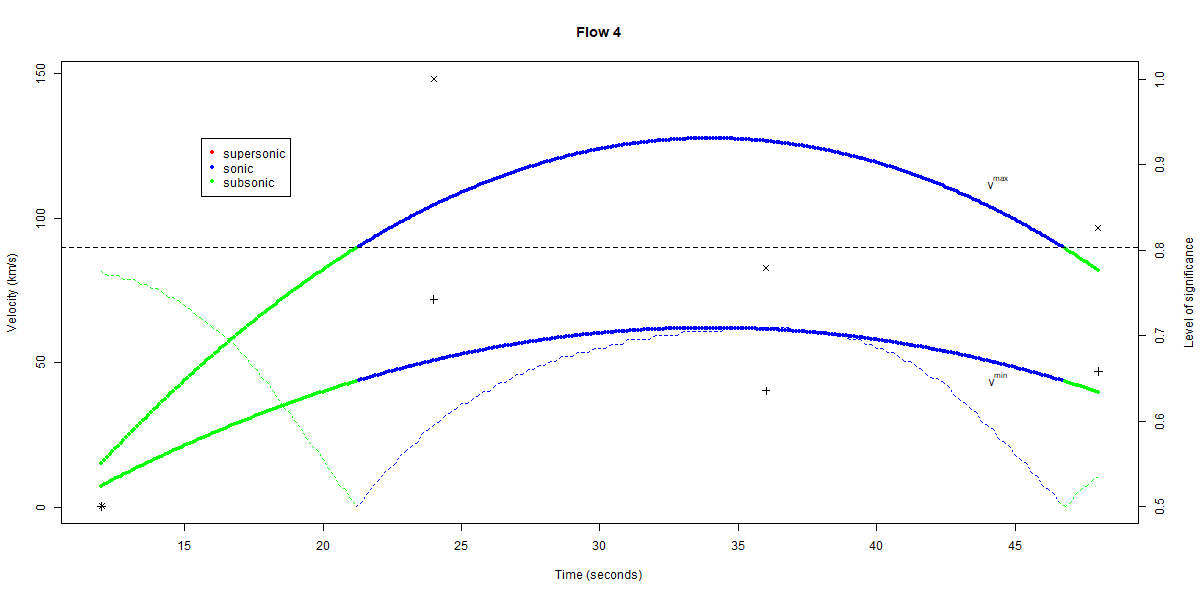}
    \includegraphics[width=6cm]{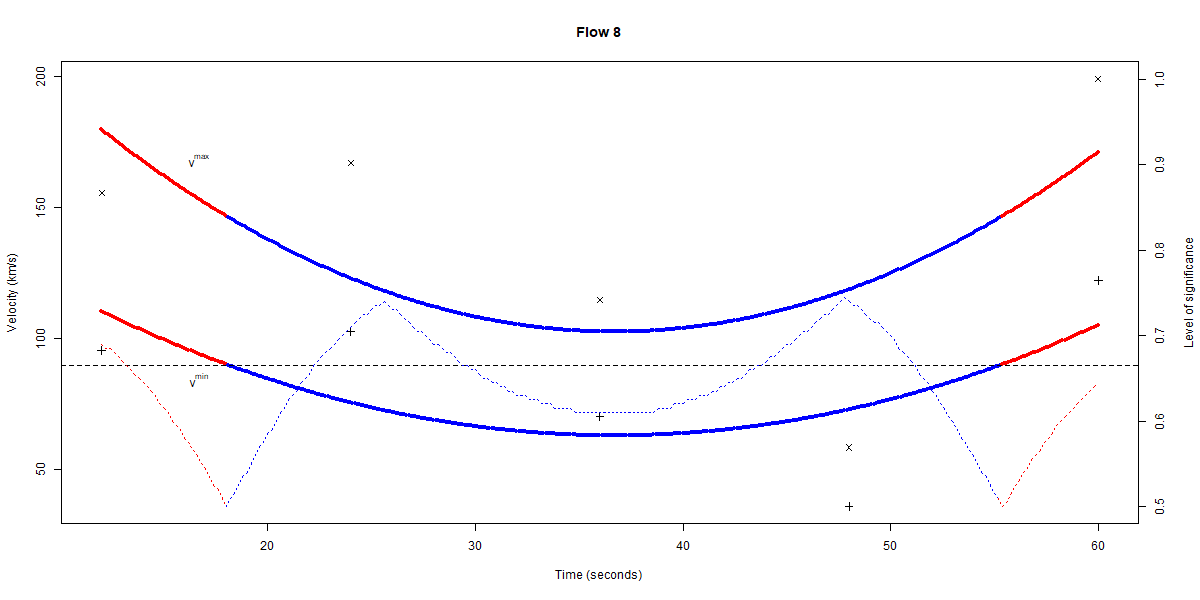}
      \includegraphics[width=6cm]{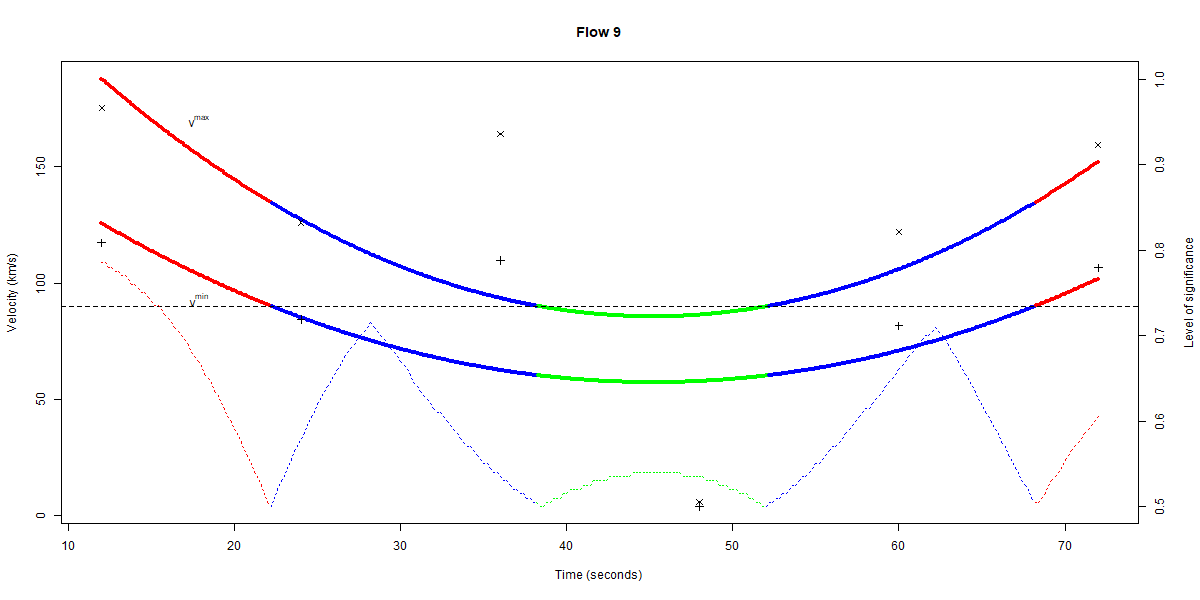}
    \includegraphics[width=6cm]{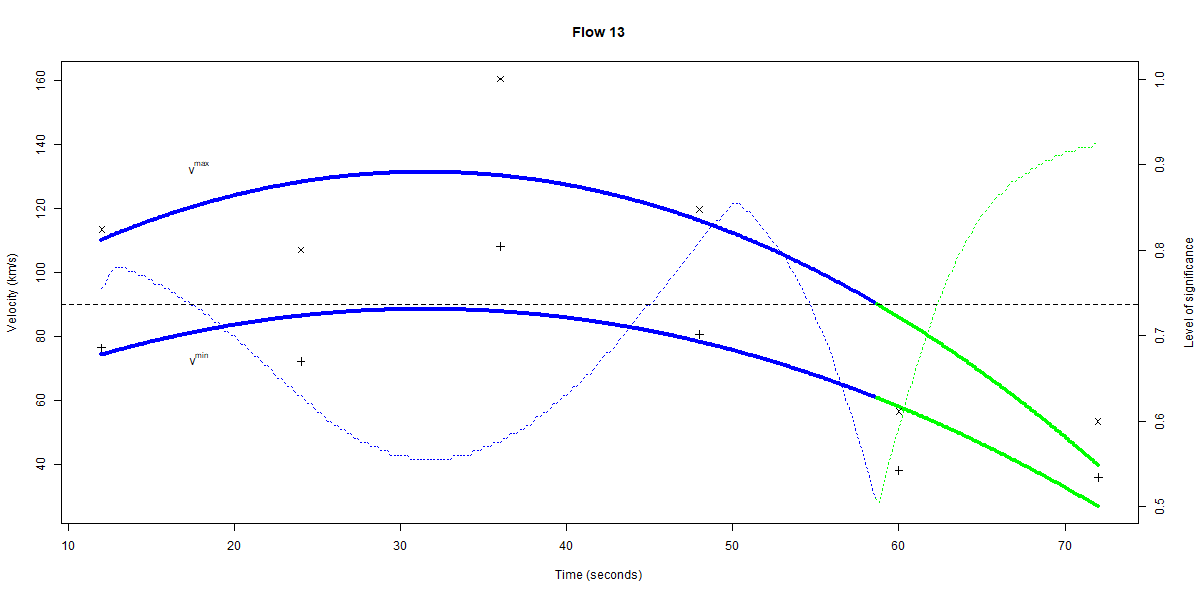}
    \includegraphics[width=6cm]{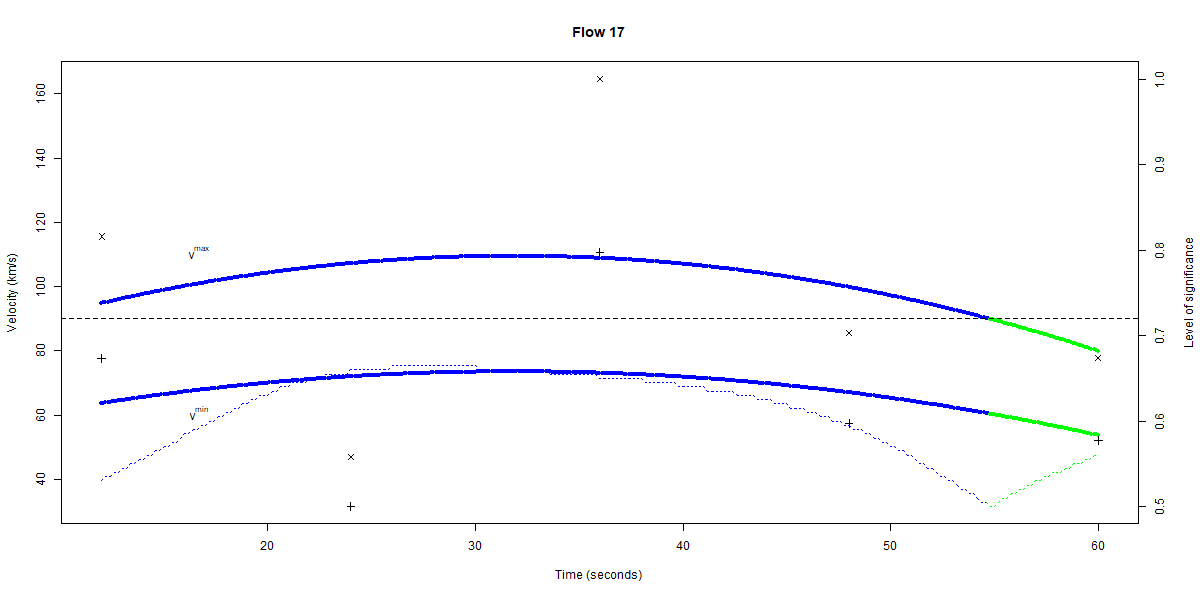}
  \caption{Same as in Fig.~\ref{supersonic} for sonic flow patterns.}
  \label{sonic}
\end{figure}

\begin{figure*}[h!]
        \includegraphics[width=6cm]{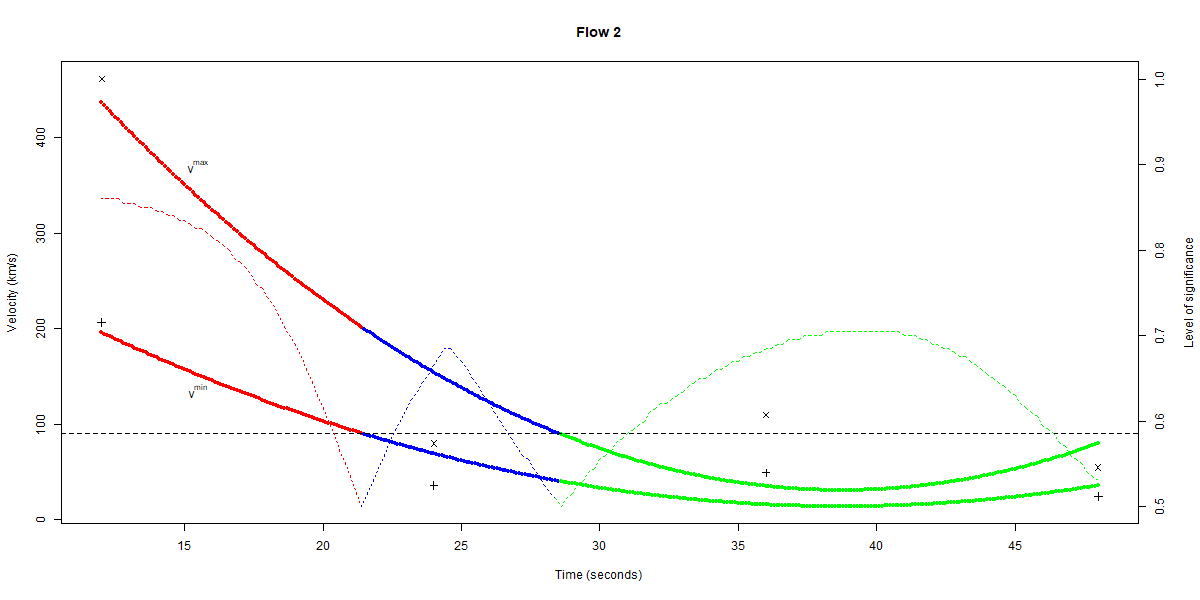}
        \includegraphics[width=6cm]{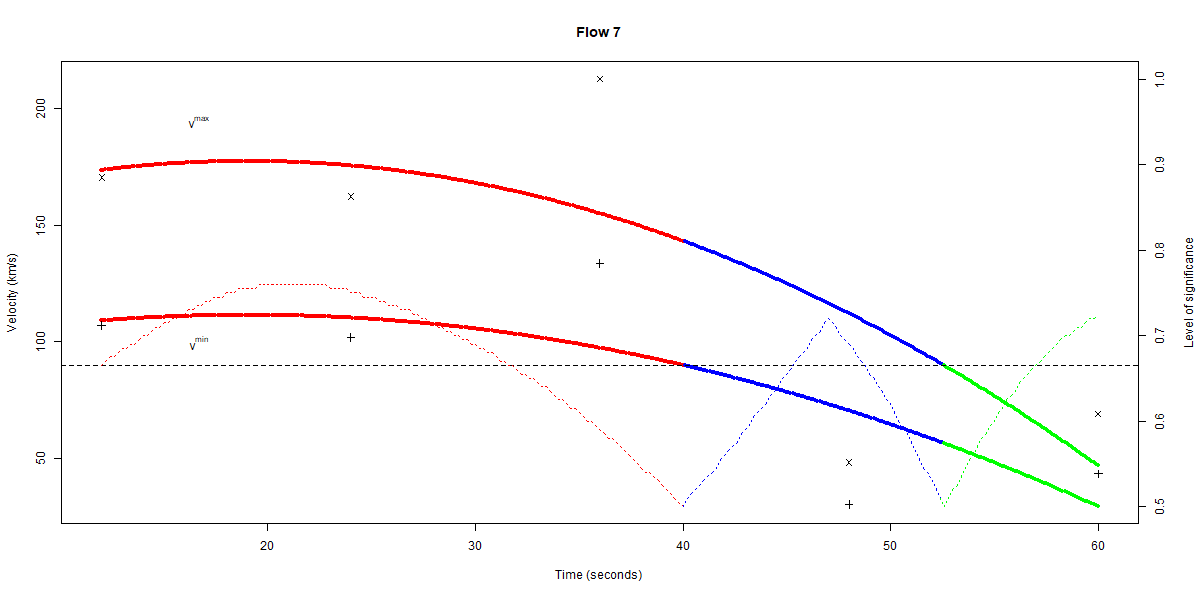}
        \includegraphics[width=6cm]{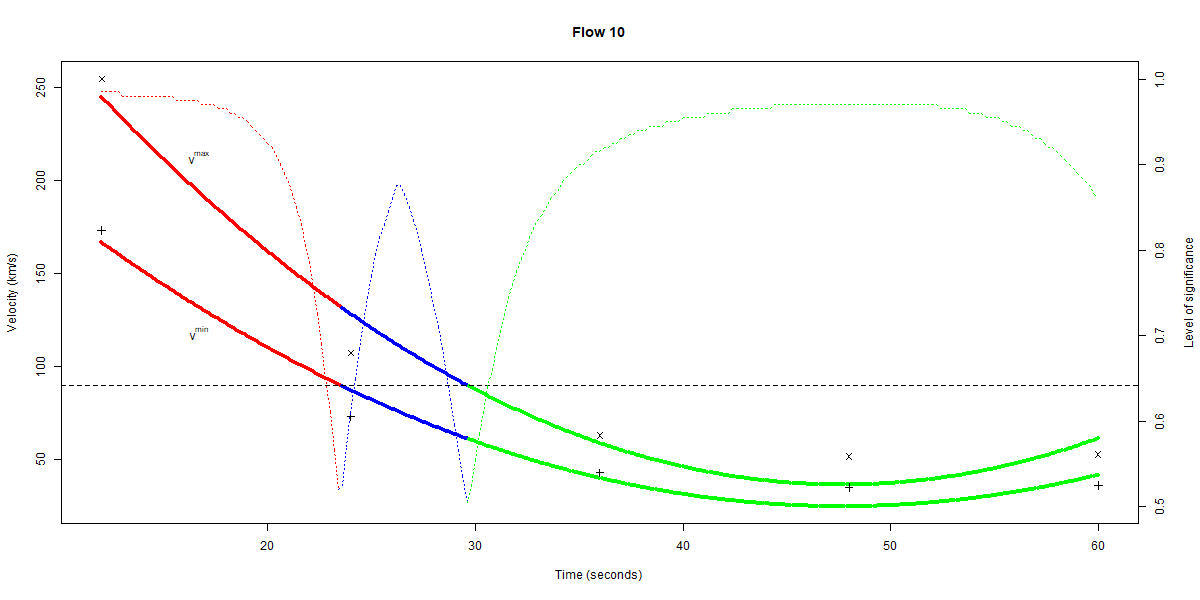}
  \caption{Same as in Fig.~\ref{supersonic} for transonic flow patterns.}
  \label{transonic}
\end{figure*}


In this paper we propose a three-stage process by which plasma flows are identified, their background temperatures and sound speeds are estimated, and their velocities are classified as either supersonic, subsonic, or sonic. We develop two novel methodologies. The first identifies plasma flows by means of a converging centroid algorithm and the second classifies their velocities as supersonic, subsonic, or sonic by means of statistical inference. This yields a total of eighteen identified plasma flows, 5 of which occurred in the limb data and 13 of which occurred in the central data.  

The parameters values $v_{min}=40km/s, v_{max}=200km/s, $  and $\Theta=\pi/6$ ensure that the  identified flows move in a consistent direction with velocities between $40km/s$ and $200km/s$. 
These parameter values can be adjusted based on the user's requirements. The smaller $v_{min}$ and the larger $v_{max}$ and $\Theta$ are chosen, the larger the number of identified flows, and vice versa.
The identified flows have a duration of around one minute and they cover a distance in the range between 4~000 and 10~000 kilometers. Out of the 18 identified plasma flows, 11 are classified as predominantly supersonic, 3 as subsonic and 3 as sonic at a $70\%$ level of confidence.
Two of these flows have large supersonic and subsonic durations, which implies that these flows form a subclass of transonic flows. 
A lower number of flows are classified at higher levels of significance, since the confidence in the classifications of these flows is higher. At a $80\%$ level of confidence, 12 flows are classified, 8 as supersonic and 4 as subsonic, and at a $90\%$ level of confidence only 6 flows are classified, 3 as supersonic and 3 as subsonic. It is important to note that 5 of the 6 flows that are classified at a $90\%$ level of significance have low standard errors (see Table~\ref{flowtable}) with the exception of Flow 11, which has an estimated velocity lower bound that is much larger than the background sound speed (see Fig.~\ref{supersonic}). This indicates that a low standard error contributes to the confidence by which a flow is classified, and vice versa. 

We utilize the upper and lower bounds of the flow velocities (upper and lower thick solid lines in each panel of Figs. 4 to 7; the velocity values are shown on the left ordinate axis) together with their corresponding significance parameter (dotted lines in the same panels, the values shown on right ordinate axis) in order to classify a flow as supersonic, subsonic, or sonic during the flow time span at a certain level of significance.
If the flow is classified as supersonic for the majority of its duration at a certain level of significance, then the flow is said to be predominantly supersonic at that level of significance, similarly for sonic or subsonic. The flows are clearly identified as supersonic, subsonic, or sonic and are illustrated in Figs.~\ref{supersonic}, \ref{subsonic}, and \ref{sonic}, respectively. In two specific cases, which we show in Fig.~\ref{transonic}, there is no clear predominant classification of the flow at a $70\%$ level of significance as the supersonic and subsonic cumulative durations are both large, although one of them dominates (in Table~\ref{flowtable70} the corresponding classification indicators are marked with an asterisk $*$). Consequently these flows transition from subsonic to supersonic states or vice versa. In such a case, the flow is said to be transonic. 
  
The method of flow identification and classification proposed in this work can be applied to SDO images of active regions in all available EUV spectral lines to reveal the characteristics of the small-scale flow patterns at different altitudes in the solar corona. For 171$\AA$ images it enables us to differentiate flows with regard to\ their velocities and their relation to the background sound speed at different levels of statistical confidence. 
In many cases we observe that the flows accelerate or decelerate and in some cases transition from one classification to another. The velocities are estimated taking into account projection due to the sphericity of the solar surface and, consequently, we determine their upper and lower bounds $v^{max}$ and $v^{min}$. We validate the applied distance correction method by matching the results of the velocity estimation close to the limb (flows 1-5) and at the visible central meridian (flows 6-18), which appeared to be comparable to each other. A further specification of the velocity curves can be achieved by evaluation of the vector magnetic field projection related uncertainties addressed in \citet{Falconer2016} and references therein and also reconstruction of the 3D structure of magnetic loops \citep{Tiwari2017}, and taking into account the orientation of different parts of the magnetic loops (presumably playing the role of the flow guides) with regard to\ the line of sight. 
This kind of modeling is beyond the scope of this study. However, building on the current results, one could make some intuitive indications. Plausibly, the effect of the magnetic loop curvature and its projection on the line of sight are mainly manifested in reduced apparent velocities compared to the real motion. This means that the method presented in this paper correctly classifies the supersonic flow patterns, while some of the sonic flows found might also be supersonic as their velocities can be underestimated because of the just-mentioned 3D field orientation. On the other hand, the proper analytical and numerical modeling of sources of information and the dynamics of intensity enhancement in magnetic loops, would shed light on the nature of the observed small-scale motions, perhaps attaching to them the role of tracers or 'fingerprints' of the small-scale magnetic structure of solar active regions. \\
  
The proposed method can be applied to any solar image data and is expected to perform better on higher resolution images such as Hi-C 2.1 data. Furthermore, the converging centroid algorithm of Sect. 2.1 can easily be extended so that flows can be located and tracked on multiple spectral lines  simultaneously. This can be done by substituting $\lambda \left( I_k(x,y) -\gamma \, I_{k-1}(x,y)\right)$ with $\sum_f \lambda_{f} \left( I^{f}_k(x,y) -\gamma \, I_{k-1}^f(x,y)\right)$ , where $f$ indicates the spectral lines of interest. It is important to note that for each additional spectral line $f$ it is necessary to specify an additional parameter value $\lambda_f$. This extension of the flow location and tracking methodology, together with further applications to high resolution solar images, such as Hi-C 2.1 data, is the matter of our forthcoming investigations.

\section{Conclusions}
The number and characteristics of plasma flows occurring on the surface of the Sun has an inherent effect on the physical conditions of the solar atmosphere. The results of the proposed methodology can be used to characterize the physical properties of the solar atmosphere, and in turn be used to analyze the cause-effect relationship between plasma flows and magnetic reconnection and energy release events. The spacial configuration of the flow presumably follows the complex structure of active regions, which recently have been confirmed and reported by high-resolution observations by \cite{Williams2020}. It is known that the distribution of characteristic sizes of magnetic strands follows a certain pattern of scale invariance with noticeable slopes of power laws in the logarithmic scale. This manifests the fact that these distributions are the result of strongly non-equilibrium statistical processes \citep{Maes2009, Maes2007} of magnetic strand formation and diffusion, which play the role of 'seeds' in coronal send-pile models that predict the presence of solar flares as avalanches of small-scale magnetic reconnection processes \citep[see][and references therein]{Baiesi2008}. This statistical method of flow velocity evaluation in EUV images can be of more wide use. For instance, coronal rains or other motions of the intensity enhancements, such as quiet Sun loops, filament channels \citep{Panesar2020}, other unstable coronal structures \citep{Bagashvili2018, Joshi2020}, or coronal rains \citep{Vashalomidze2015, Antoli2015} can also be  traced in a similar manner; only the centroid convergence and the following parameters need to be adjusted accordingly. In the future, such an advanced statistical analysis would be useful for identifying the characteristics of plasma flows as inputs to a machine learning or neural network algorithms, where the result would be a characterisation of an active region and the measurement of its evolution.

\vspace{10mm}

\begin{acknowledgements}
The work was supported by Shota Rustaveli National Science Foundation grants
FR17\_609, DI-2016-52.  Work of E.P was supported under Shota Rustaveli National Science Foundation grant for young scientists for scientific research internships abroad IG/46/1/16. Work of B.M.S. was supported by the Austrian Fonds zur Forderung der wissenschaftlichen Forschung (FWF) under projects P25640-N27,
S11606-N16. S.P. was supported by the projects C14/19/089  (KU Leuven), G.0D07.19N  (FWO-Vlaanderen), C 90347 (ESA Prodex).
M.L.K. additionally acknowledges support from the projects I2939-N27 and S11606-N16 of the Austrian Science Fund (FWF) and grant number 075-15-2019-1875 from the government of the Russian Federation under the project called “Study of stars with exoplanets”. We are thankful to the anonymous reviewer for the valuable remarks that led to the significant improvement of the manuscript content.
\end{acknowledgements}

\bibliographystyle{aa}
\bibliography{biblio}

\end{document}